\documentclass[conference]{IEEEtran}
\usepackage[letterpaper, top=0.7in, bottom=1in, left=0.58in, right=0.58in]{geometry}

\usepackage{blindtext, graphicx}
\newcommand\blfootnote[1]{%
  \begingroup
  \renewcommand\thefootnote{}\footnote{#1}%
  \addtocounter{footnote}{-1}%
  \endgroup
}

\usepackage{authblk}
\usepackage{balance}  
\usepackage{fixltx2e}
\usepackage{algorithmicx}
\usepackage{algorithm}
\usepackage[noend]{algpseudocode}
\usepackage{url}
\let\labelindent\relax
\usepackage{enumitem}
\usepackage{multirow}
\usepackage{graphicx}
\usepackage{balance}  
\usepackage{url}
\usepackage[acronym,toc,shortcuts]{glossaries}
\usepackage{epsfig}
\usepackage{amssymb}
\usepackage{amsthm}
\usepackage{multirow}
\usepackage{caption}
\usepackage{subcaption}
\usepackage{siunitx}
\usepackage{soul}
\usepackage{algorithm}
\usepackage{algpseudocode}
\usepackage{csquotes}
\usepackage{bbm}
\usepackage{dsfont}
\usepackage{enumitem}
\usepackage{adjustbox}
\usepackage{array}
\usepackage{multirow}
\usepackage{fancyhdr}
\usepackage[T1]{fontenc}
\usepackage{balance}
\usepackage[table]{xcolor}
\usepackage{comment}

\usepackage[acronym,toc,shortcuts]{glossaries}

\graphicspath{{figures/}}

\newacronym{3GPP}{3GPP}{The 3rd Generation Partnership Project }
\newacronym{5G}{5G}{Fifth Generation}
\newacronym{AAA}{AAA}{Authentication, Authorization and Accounting}
\newacronym{AI}{AI}{Artificial Intelligence}
\newacronym{AES}{AES}{Advanced Encryption System}
\newacronym{AoI}{AoI}{Age of information}
\newacronym{AP}{AP}{Access Point}
\newacronym{API}{API}{Application Programming Interface}
\newacronym{APN}{APN}{Access Point Name}
\newacronym{BS}{BS}{Base Station}
\newacronym{BER}{BER}{Bit Error Rate}
\newacronym{BSSID}{BSSID}{Basic Service Set Identification}
\newacronym{CAT}{CAT}{Capacity-Aware TOPSIS}
\newacronym{CEP}{CEP}{Complex Event Processing}
\newacronym{CELL-ID}{CELL-ID}{cell identification ID}
\newacronym{CGI}{CGI}{Cell Global Identification}
\newacronym{CI}{CI}{Confidence Interval}
\newacronym{CLSM}{CLSM}{Closed loop spatial multiplexing}
\newacronym{CQI}{CQI}{Channel Quality Indicator}
\newacronym{CN}{CN}{core network}
\newacronym{CNN}{CNN}{Convolutional Neural Networks}
\newacronym{CL}{CL}{Closed-Loop}
\newacronym{CoMP}{CoMP}{coordinated multi-point}
\newacronym{CPU}{CPU}{Central Processing Unit}
\newacronym{CS}{CS}{central scheduler}
\newacronym{CSI}{CSI}{Channel Status Information}
\newacronym{eNB}{eNB}{evolved Node-B}
\newacronym{DaaS}{DaaS}{Data as a Service}
\newacronym{DL}{DL}{Downlink}
\newacronym{DLT}{DLT}{Distributed Ledger Technology}
\newacronym{DMM}{DMM}{Distributed Mobility Management}
\newacronym{ECA}{ECA}{Event-Condition-Action}
\newacronym{ECC}{ECC}{Elliptic-curve cryptography}
\newacronym{eNodeB}{eNodeB}{evolved Node-B}
\newacronym{E-RAB}{E-RAB}{E-UTRAN Radio Access Bearer}
\newacronym{ETSI}{ETSI}{European Telecommunications Standards Institute}
\newacronym{FDD}{FDD}{Frequency Division Duplexing }
\newacronym{FEM}{FEM}{Flow Extraction Manager}
\newacronym{GDPR}{GDPR}{General Data Protection Regulation}
\newacronym{GGSN}{GGSN}{Gateway GPRS Support Node}
\newacronym{GPRS}{GPRS}{General packet radio service}
\newacronym{GTP}{GTP}{GPRS Tunneling Protocol}
\newacronym{HetNet}{HetNet}{heterogeneous network}
\newacronym{HSS}{HSS}{Home Subscriber Station}
\newacronym{HTTP}{HTTP}{Hypertext Transfer Protocol}
\newacronym{HDFS}{HDFS}{Hadoop Distributed File System}
\newacronym{HiveQL}{HiveQL}{Hive Query language}
\newacronym{HSPA}{HSPA}{High Speed Packet Access}
\newacronym{IBLER}{IBLER}{Initial Block Error Rate}
\newacronym{ICIC}{ICIC}{inter-cell interference coordination}
\newacronym{ICN}{ICN}{information-centric network}
\newacronym{IEEE}{IEEE}{Institute of Electrical and Electronics Engineers}
\newacronym{IETF}{IETF}{Internet Engineering Task Force}
\newacronym{IMSI}{IMSI}{International Mobile Subscriber Identity}
\newacronym{IMEI}{IMEI}{International Mobile Station Equipment Identity}
\newacronym{IMS}{IMS}{IP Multimedia Subsystem}
\newacronym{ICMP}{ICMP}{Internet Control Message Protocol}
\newacronym{IoT}{IoT}{Internet of Things}
\newacronym{InP}{InP}{Infrastructure Provider}
\newacronym{ITU}{ITU}{International Telecommunication Union}
\newacronym{IT}{IT}{Information Technology}
\newacronym{GBR}{GBR}{Guaranteed Bit Rate}
\newacronym{GLUE}{GLUE}{General Language Understanding Evaluation}
\newacronym{JSON}{JSON}{JavaScript Object Notation}
\newacronym{KPI}{KPI}{Key Performance Indicator}
\newacronym{LA}{LA}{Location Area}
\newacronym{LAC}{LAC}{location area code}
\newacronym{LMA}{LMA}{Local Mobility Anchor}
\newacronym{LTE}{LTE}{long term evolution}
\newacronym{MADM}{MADM}{Multiple Attribute Decision Making}
\newacronym{MCC}{MCC}{Mobile Country Code}
\newacronym{MCS}{MCS}{Modulation Coding Scheme}
\newacronym{MNC}{MNC}{Mobile Network Code}
\newacronym{MIMO}{MIMO}{multiple-input multiple-output}
\newacronym{MAG}{MAG}{Mobile Access Gateway}
\newacronym{MAAR}{MAAR}{Mobility Anchor and Access Router}
\newacronym{ML}{ML}{Machine Learning}
\newacronym{MME}{MME}{Mobility Management Entity}
\newacronym{MN}{MN}{Mobile Node}
\newacronym{MNO}{MNO}{Mobile Network Operator}
\newacronym{MSISDN}{MSISDN}{Mobile Station International Subscriber Directory Number}
\newacronym{NBI}{NBI}{NorthBound Interface}
\newacronym{NIST}{NIST}{National  Institute  of  Standards  and Technology}
\newacronym{NLP}{NLP}{Natural Language Processing}
\newacronym{NLU}{NLU}{Natural Language Understanding}
\newacronym{NOMA}{NOMA}{Non-Orthogonal Multiple Access}
\newacronym{NoSQL}{NoSQL}{Not Only SQL}
\newacronym{NR}{NR}{New Radio}
\newacronym{QoS}{QoS}{quality-of-service}
\newacronym{QoE}{QoE}{quality-of-experience}
\newacronym{OAM}{OAM}{Operation, Administration and Management}
\newacronym{ONF}{ONF}{Open Networking Foundation}
\newacronym{ONOS}{ONOS}{Open Network Operating System}
\newacronym{OS}{OS}{operating system}
\newacronym{OL}{OL}{Open-Loop}
\newacronym{PDN}{PDN}{packet data network}
\newacronym{PF}{PF}{Proportional Fair}
\newacronym{P-GW}{P-GW}{packet gateway}
\newacronym{PDP}{PDP}{Packet Data Protocol}
\newacronym{PHY}{PHY}{physical layer}
\newacronym{PMIPv6}{PMIPv6}{Proxy Mobile IPv6}
\newacronym{PMI}{PMI}{Precoding Matrix Index}
\newacronym{PoW}{PoW}{Proof-of-Work}
\newacronym{PQC}{PQC}{Post-Quantum Cryptography}
\newacronym{PRB}{PRB}{Physical Resource Block}
\newacronym{PUSCH}{PUSCH}{Physical Uplink Shared Channel}
\newacronym{QAM}{QAM}{Quadrature amplitude modulation}
\newacronym{QRSA}{QRSA}{Quantum Resistant Security Algorithm}
\newacronym{QCI}{QCI}{QoS Class Identifier}
\newacronym{RA}{RA}{Routing Area}
\newacronym{RB}{RB}{Resource Block}
\newacronym{RI}{RI}{Rank Indicator}
\newacronym{RAN}{RAN}{radio access network}
\newacronym{RFC}{RFC}{Request for Comment}
\newacronym{RRC}{RRC}{Radio Resource Control}
\newacronym{RNC}{RNC}{radio network controller}
\newacronym{RNN}{RNN}{Recurrent Neural Networks}
\newacronym{RSA}{RSA}{Rivest–Shamir–Adleman}
\newacronym{RSSI}{RSSI}{Received Signal Strength Indicator}
\newacronym{RSRP}{RSRP}{Reference Signal Received Power}
\newacronym{OTT}{OTT}{over-the-top}
\newacronym{SA}{SA}{Stand Alone}
\newacronym{SAC}{SAC}{service area code}
\newacronym{SCMA}{SCMA}{Sparse Code Multiple Access}
\newacronym{SLA}{SLA}{Service Level Agreement }
\newacronym{SDN}{SDN}{Software Defined Networking}
\newacronym{SDO}{SDO}{Standards Developing Organization}
\newacronym{SFN}{SFN}{Single Frequency Network}
\newacronym{S-GW}{S-GW}{serving gateway}
\newacronym{SINR}{SINR}{signal-to-interference-plus-noise ratio}
\newacronym{SGSN}{SGSN}{Serving GPRS Support Node}
\newacronym{SP}{SP}{Service Provider}
\newacronym{SSID}{SSID}{Service Set Identification}
\newacronym{SVD}{SVD}{singular value decomposition}
\newacronym{TCP}{TCP}{transport control protocol}
\newacronym{TDD}{TDD}{Time Division Duplexing}
\newacronym{TM}{TM}{transmission mode}
\newacronym{TEID}{TEID}{tunnel endpoint identifier}
\newacronym{UDN}{UDN}{Ultra Dense Network}
\newacronym{UMTS}{UMTS}{Universal Mobile Telecommunications Service} 
\newacronym{UE}{UE}{user equipment}
\newacronym{UL}{UL}{Uplink}
\newacronym{UDP}{UDP}{User Datagram Protocol}
\newacronym{VM}{VM}{Virtual Machine}
\newacronym{VNF}{VNF}{Virtual Network Function}
\newacronym{WiFi}{WiFi}{Wireless Fidelity}
\newacronym{WLAN}{WLAN}{Wireless Local Area Network}

\newacronym{DDoS}{DDoS}{Distributed Denial-of-Service}
\newacronym{DNS}{DNS}{Domain Name System}
\newacronym{DNSSEC}{DNSSEC}{DNS Secure}
\newacronym{DoC}{DoC}{DNS over CoAP}
\newacronym{DoE}{DoE}{DNS-over-Encryption}
\newacronym{DoH}{DoH}{DNS over Hypertext Transfer Protocol Secure (HTTPS)}
\newacronym{DoQ}{DoQ}{DNS over QUIC}
\newacronym{DoT}{DoT}{DNS over TLS}
\newacronym{TLS}{TLS}{Transport Layer Security}

\begin{document}
%
\title{Analysis of Robust and Secure DNS Protocols for IoT Devices}

\author{Abdullah Aydeger$^{\diamond}$, Sanzida Hoque$^{\diamond}$, Engin Zeydan$^{\ast}$,  Kapal Dev$^{\vartriangle}$\\
$^{\diamond} $Florida Institute of Technology, Melbourne, FL, USA, 32901\\
$^{\ast} $Centre Tecnològic de Telecomunicacions de Catalunya (CTTC), Castelldefels, Barcelona, Spain, 08860.\\
$^{\vartriangle}$ Munster Technological University, Bishopstown, Cork, Ireland.\\
\protect Emails: aaydeger@fit.edu, shoque2023@my.fit.edu engin.zeydan@cttc.cat, kapal.dev@ieee.org}


\maketitle

\begin{abstract}
The DNS (Domain Name System) protocol has been in use since the early days of the Internet. Although DNS as a de facto networking protocol had no security considerations in its early years, there have been many security enhancements, such as DNSSec (Domain Name System Security Extensions), DoT (DNS over Transport Layer Security), DoH (DNS over HTTPS) and DoQ (DNS over QUIC). With all these security improvements, it is not yet clear what resource-constrained Internet-of-Things (IoT) devices should be used for robustness. In this paper, we investigate different DNS security approaches using an edge DNS resolver implemented as a Virtual Network Function (VNF) to replicate the impact of the protocol from an IoT perspective and compare their performances under different conditions. We present our results for cache-based and non-cached responses and evaluate the corresponding security benefits. Our results and framework can greatly help consumers, manufacturers, and the research community decide and implement their DNS protocols depending on the given dynamic network conditions and enable robust Internet access via DNS for different devices.

\end{abstract}


\IEEEpeerreviewmaketitle


\section{Introduction}
\ac{DNS} is a networking protocol that maps human-readable domain names (e.g., www.mywebsite.com) to IP addresses (e.g., 192.168.0.1). It works via a hierarchical structure of servers, whereby each server is responsible for a portion of the domain namespace. When a client wants to resolve a domain name (i.e., a website's well-known name), it sends a query to a DNS resolver, which then contacts authoritative servers from top to bottom until it finds the IP address for the requested domain name. Traditionally, the DNS resolver can be hosted and operated by an Internet Service Provider, a mobile carrier, a WIFI network, or another third party. In recent years, the \ac{DNS} protocol has received much attention from various organizations and academia due to the lack of encryption or integrity of traditional DNS protocol messages \cite{hanna2023performance, harrilal2023bringing}. In order to improve security, some of the protocols have been proposed and are currently used by some DNS servers: \ac{DNSSEC}, DNS over \ac{TLS}, \ac{DoH}, and \ac{DoQ}. In short, \ac{DNSSEC} adds a layer of security to the DNS by digitally signing data to ensure its authenticity and integrity, preventing attacks like DNS spoofing. \ac{DoT} encrypts DNS queries and responses between the client (e.g., your computer) and the DNS resolver using TLS, protecting your privacy and preventing eavesdropping. \ac{DoH} also encrypts DNS traffic but uses the HTTPS protocol, making it more difficult for network administrators to block or filter DNS requests. \ac{DoQ} is a newer protocol that combines the encryption of \ac{DoT} with the speed and efficiency of QUIC, a modern transport protocol designed for faster and more robust internet connections. It is important to note that DNS security protocols have their pros and cons and vary in performance depending on network conditions and available hardware resources. 
\blfootnote{This paper has been accepted in the 2025 IEEE International Conference on Communications (ICC): SAC Cloud Computing, Networking, and Storage Track. The copyright belongs to IEEE, and the final version will be published in the IEEE Xplore.}
While it is a challenge to set up robust security protocols for resource-constrained devices when they interact with the DNS \cite{9133283}, \cite{aucklah2021impact}, these devices are also more vulnerable to DNS-based attacks such as cache poisoning, man-in-the-middle attacks, DNS tunneling and other which impacts their robust operations. A large number of IoT devices collect and transmit user data. Without encryption and authentication, this data is vulnerable to interception and exploitation by attackers. In addition, IoT devices are often interconnected and an integral part of a larger ecosystem. If a single compromised device interacts with the DNS infrastructure, this can lead to the spread of attacks to the entire network. According to a report by Forescout Research Labs, over 100 million connected resource-constrained devices could be vulnerable to nine recently disclosed security flaws discovered in the DNS implementations known as Name:Wreck \cite{dos2021name}.
Therefore, it is important to explore robust and secure DNS protocols for resource-constrained devices. The traditional DNS protocol and infrastructure have been very attractive targets for cyberattacks in the past because they have many vulnerabilities that could be exploited. Although secure DNS protocols allow the encryption of DNS traffic, some recent studies have presented attack models against these protocols. Specifically, \cite{hynek2022summary} gives an overview of the current state of research on \ac{DoH} and various malicious and unwanted activities that exploit this technology. They recommend further investigation into the web-based abuse of \ac{DoH} \cite{hynek2022summary}. The paper in \cite{trevisan2023attacking} conducted evaluations with passive and active attacks to investigate the effectiveness of \ac{DoH} and Encrypted Client Hello (ECH) in preserving user privacy. The authors found that passive attacks were very effective in extracting encrypted domain names. As some general approaches to securing DNS attacks, researchers have proposed DNS as a \ac{VNF}, and the use of some moving target defense-based deception techniques to protect against cyberattacks that can exploit DNS servers and protocol \cite{aydeger2024mtdns, nesary2022vdns}. In this paper, however, our main focus is on comparing performance metrics for the latest secure DNS protocols rather than analyzing the security of the protocols themselves. 


Recently,  evaluating and comparing different DNS security protocols and approaches have been of interest. The authors in \cite{lu2019end} present a large-scale analysis of \ac{DoE} using data collected from various sources. In \cite{bottger2019empirical}, the authors examine DoH, focusing on the assessment of the costs associated with the additional security. They conclude that the use of DoH can lead to performance degradation and increased latency. In a study conducted by \cite{hounsel2020comparing}, DoH, DoT, and traditional DNS are analyzed under different network conditions to evaluate how they affect web performance. Using Cloudflare's DNS resolver, high-speed and simulated mobile networks with varying degrees of latency and loss were measured.
On the other hand, DoC (DNS over CoAP)~\cite{lenders2023securing} is a newly proposed DNS resolution protocol that is specifically designed for resource-constrained devices. Ultimately, the choice of protocol depends on the specific characteristics of the network used and the balance between response time and page load time. The researchers in \cite{chhabra2021measuring} measured the performance of Google's and Cloudflare's DoH services using a customized measurement tool named "dohtest". Response time, success rates, and throughput of DoH requests were analyzed for 50 different locations around the world. In another study, the researchers conducted several experiments to investigate the efficiency of DoQ compared to conventional DNS protocols (DNS over UDP and DNS over TCP) \cite{kosek2022dns}.

\subsection{Edge DNS}

In this paper, we consider the operation of a local DNS resolver as a \ac{VNF}, which can also be used by resource-constrained devices. In other words, we consider an edge \ac{DNS} service that emulates the functionalities of consumer resource-constrained network devices. With edge \ac{DNS}, a distributed network of servers can resolve DNS queries efficiently and quickly. In this way, IoT devices receive responses from the nearest available server, reducing latency and improving performance. By managing DNS records and configurations in real time, edge DNS provides organizations with greater flexibility and control. In addition to advanced features such as traffic management, load balancing, and security measures such as \ac{DDoS} protection, edge DNS is a valuable tool for ensuring the availability and security of online services. Ultimately, edge DNS offers significant benefits for organizations looking to improve the reliability, security, and flexibility of their DNS infrastructures.

Our main focus is on comparing different secure DNS protocols when DNS run at the edge as \ac{VNF} to develop a better understanding and robust DNS for resource-constrained devices. To the best of our knowledge and from the related work we have researched, there is no work that considers and examines the performance of DNSSEC, DoT with/without locally held signature zone files. We believe that our work can provide a general idea to the industry considering providing DNS services either internally or commercially to others. We test the performance of different DNS protocols with and without security safeguards to help consumers decide whether the privacy benefits of secure DNS protocols outweigh their overhead, especially in resource-constrained environments, and offer helpful insights to the research community. 

\section{Secure DNS Protocols for Resource Constrained Devices}
\label{secureDNS}

\subsection{Secure DNS Protocols}

\begin{figure}[htp!]
\centering
\includegraphics[width=\linewidth]{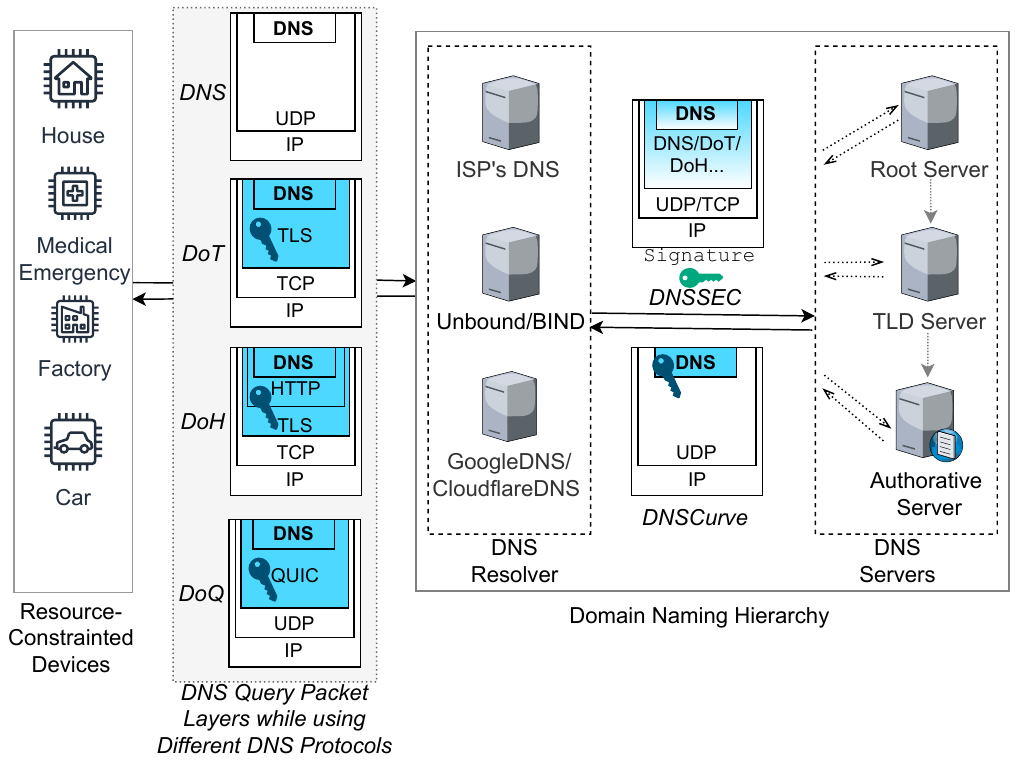}
\vspace{-0.5cm}
\caption{Illustration of various DNS Security Protocols.} 
\label{fig:dns}
\vspace{-0.5cm}
\end{figure}

Various DNS security protocols that are widely accepted and used by the community are shown in Figure \ref{fig:dns}. The blue segment in this figure indicates encrypted content \cite{lyu2022survey}. \textit{(i) DNS Security Protocol (DNSSEC):} In DNSSEC, digital signatures are used to verify the integrity and authenticity of DNS records so that users can verify the accuracy of information provided by DNS servers. \textit{(ii) \ac{DoT}:} The DoT protocol ensures privacy and integrity by encrypting DNS traffic with the same Transport Layer Security (TLS) protocol. HTTPS websites also use this for encryption and authentication. \textit{(iii) \ac{DoH}:} DoH is a replacement for DoT that encrypts DNS traffic but, unlike DoT, uses the same port as HTTPS traffic, concealing DNS traffic within other HTTPS traffic and preventing eavesdropping and data manipulation by encrypting all data between the client and the DoH-based DNS resolver using the HTTPS protocol. \textit{(iv) \ac{DoQ}:} QUIC, which was proposed by Google to improve the performance of encrypted DNS, uses a TLS-like encryption system and combines connection establishment and key agreement into a single round-trip time (RTT). Unlike DoT and DoH, DoQ uses the QUIC protocol over UDP, which enables faster and lighter encrypted communication by eliminating the initial TCP handshake.

Note that the secure version of DNS protocols has already been adopted by many applications, including 5G and IoT \cite{hesselman2020dns}. Nevertheless, configuring the DNS protocol and resolver selection according to consumer preferences (i.e., based on resource constraints) is not straightforward, and none of the work in this area has considered a solution to this problem.

\subsection{Impact of Resource-Constrains on IoT devices} 

IoT devices are characterized by limited computing resources, which include limited computing power, memory, storage, and power supply. These devices are essential for many applications but present unique challenges in terms of reliability, security, and functionality. There are several implications of these limitations. First, these devices often use low-power microcontrollers, such as 8-bit or 16-bit, with limited CPU capabilities (with an average CPU clock speed of 100-400 MHz) to keep energy consumption low. This limits their ability to perform complex computations or run intensive security protocols. Second, resource-constrained IoT devices typically have only a tiny amount of RAM and flash memory, ranging from a few kilobytes to a few megabytes, which limits the amount of data they can process and store. This affects their ability to run large software programs or store extensive logs. Thirdly, many IoT devices are battery-powered and designed to operate for long periods of time with minimal energy consumption. Energy efficiency is critical, which often means they cannot afford the energy costs of complex operations. 

Fourth, many of these devices use low-power communication systems, which naturally offer lower bandwidth to extend battery life. High bandwidth communication consumes more power, which would drain the battery of some low-power devices. Limited memory also makes it difficult to handle high-bandwidth communication, as it does not provide enough temporary storage for incoming and outgoing data packets. Typically, an IoT device can utilize a few kilobits per second (kbps) to a few megabits per second (Mbps). Normally, the memory and bandwidth requirements of the DNS are significantly higher than those of IoT devices, as it is not intended for constrained devices \cite{ayoub2023dns}. Fifth, these devices are often small and inexpensive, making them ideal for widespread use but limiting the space and budget for robust hardware security features. Sixth, in terms of connectivity, various communication protocols (e.g., Zigbee, LoRaWAN, Bluetooth Low Energy, Wi-Fi) are optimized for low power consumption but may have limitations in terms of data rate and range.


For the reasons mentioned above, limited resources make it difficult to implement traditional security measures such as strong encryption, multi-factor authentication, and regular updates. This makes them vulnerable to attacks and exploits. Due to limited memory and computing power, processing large amounts of data, performing real-time analysis, and storing historical data can be a challenge. To save energy, these devices often use low-power communication protocols, which can lead to lower data rates and higher latency. Ensuring robust communication while maintaining efficiency is one of the biggest challenges. Although IoT devices are designed to be used in large numbers, managing and updating thousands or millions of devices with limitations can be complex and resource-intensive.

\subsection{Strategies for  limitations}

Providing secure DNS for the resource-constrained IoT has many challenges to be covered: (i) Although these protocols provide security through encryption, they introduce additional overhead compared to traditional DNS over UDP. This overhead can be significant for devices with limited processing and memory capabilities. (ii) The encryption and decryption processes required for secure DNS protocols consume more energy than plain DNS. This can be a major concern for battery-powered IoT devices, where energy efficiency is critical. (ii) Implementing and maintaining secure DNS protocols on resource-constrained devices can be challenging due to their limited resources and potential incompatibility with existing software stacks. (iv) Some networks or firewalls may block or interfere with encrypted DNS traffic, causing connectivity issues for IoT devices that rely on DNSSEC, DoH or DoT.

Strategies for dealing with resource constraints include developing lightweight software that is optimized for performance on limited hardware. This includes the following: (i) The use of efficient coding techniques and lightweight algorithms. Energy harvesting technologies (e.g., solar power) can be used to extend battery life and reduce energy constraints. (ii) Offloading intensive processing tasks to more powerful edge devices or gateways. This reduces the load on individual IoT devices and utilizes more powerful local resources. (iii) The use of protocols specifically designed for low-power and constrained environments, such as CoAP (Constrained Application Protocol) and MQTT (Message Queuing Telemetry Transport). (iv) Implementation of mechanisms for regular security updates that are small and can be applied in an energy-efficient manner. (v) The use of security measures that adapt to the current resource availability of the device to ensure security without compromising functionality.

\section{Experimental Analysis}

In this section, we describe our experimental environment and the metrics used to compare DNS security protocols and discuss the results. Our experiments aim to investigate the different metrics involved in securing DNS servers and messages by comparing different DNS setups, including traditional DNS (without security), DNSSEC, DoT, DoH, and DoT+DNSSEC (i.e., DoT with the signature fields for the DNS response information). This will allow us to better understand these setups and determine the most efficient approach to \ac{VNF}-based DNS (i.e., DNS from the edge perspective) that can be deployed in a given scenario. In our experimental evaluations, we consider the following metrics to compare different DNS security protocols: \textit{(i) Network Traffic:} This metric measures the total amount of network traffic in bytes for each DNS protocol packet exchange from the initial request to the receipt of the response. \textit{(ii) Response Time:} This metric for obtaining the IP address of the domain server name is the time it takes to get the response from the recursive DNS server. With this metric, we observe the time for both local and remote domain names where we use different authoritative DNS resolvers.


We have chosen BIND9 as the DNS resolver implementation. When configuring BIND9, we include external DNS servers as forwarders in case the queried domains are not contained in the local zone file or in the cache. The server forwards the query to the listed servers, where they are checked for corresponding records. In our experiment, we used three different forwarders (i.e., Google, Cloudflare, and OpenDNS) to observe how much overhead is incurred when forwarding requests compared to local responses. For the domain names, we have used different regions, such as the UK, Australia, South Africa, Ecuador, and China, as well as the data from the local zone files. Our setup is located in the USA. DNSPerf was used because of its practical features that allow large amounts of queries to be sent in the shortest possible time and an automatic calculation of the minimum, maximum, and average latency for the DNS responses. 
Several tests were carried out, and the average values for the latency in milliseconds before and after caching were calculated. We also captured the network traffic to determine the differences in packet sizes (in bytes) for each of the above protocol cases.

\begin{table*}[htp!]
\begin{center}
\scriptsize
\caption{Average Latency with Performance Loss and Formulated Security Benefits}
\begin{tabular}{ | m{2.0cm} | m{2.4cm} | m{2.9cm} | m{8cm} | m{1.5cm} | } 
  \hline
  \rowcolor{gray!30}
  DNS Protocol & Average Latency after Cache (ms) & Average Latency Performance Loss (\%) & Security Benefits Description & Security Score (0-4) \\ 
  \hline\hline
  DNS & 0.103 & N.A.\% & No encryption, authentication, or integrity; vulnerable to attacks like spoofing and cache poisoning. & 0 \\ 
  \hline
  DNSSEC & 0.199 & 93.2\% & Provides authentication and integrity through digital signatures; prevents data tampering. & 2 \\ 
  \hline
  DoT & 0.600 & 482.5\% & Encrypts DNS queries and responses via TLS; provides confidentiality, authentication, and integrity. & 3 \\ 
  \hline
  DoT + DNSSEC & 0.657 & 537.9\% & Combines DoT's encryption with DNSSEC's integrity and authentication, offering robust protection. & 3.5 \\ 
  \hline
  DoH & 0.728 & 606.8\% & Provides encryption, authentication, and integrity over HTTPS, with additional privacy through traffic obfuscation. & 4 \\ 
  \hline
\end{tabular}
\label{tab:latency_performance_security}
\end{center}
\end{table*}

\subsection{Results}

\subsubsection{\textbf{Security Quantification of DNS Protocols}}

We quantify the security benefits in Table \ref{tab:latency_performance_security} by assigning points to each of the following security properties for each DNS protocol: (i) \textit{Encryption:} Protects the confidentiality of data by ensuring that DNS queries and responses cannot be read by unauthorized parties. No encryption is awarded 0 points and if it is present, 1 point is awarded. \textit{(ii) Authentication:} Ensures that the data comes from a legitimate source and has not been tampered with. No authentication is awarded 0 points and if it is present, 1 point is awarded. \textit{(iii) Integrity:} Protects against tampering by ensuring that the data has not been altered during transit. No integrity is awarded 0 points and if it is present, 1 point is awarded.  \textit{(iv) Advanced Privacy Protection:} Additional mechanisms like traffic obfuscation or blending DNS traffic with other types of traffic to enhance privacy. No advanced privacy protection is awarded 0 points and if it is present, 1 point is awarded. The total security benefit for a protocol is calculated as summation of above four properties.

\textit{DNS (i.e., over UDP)} offers no encryption, i.e. the DNS queries and responses are sent in plaintext and can easily be intercepted by unauthorized parties. There is also no authentication, so there is no guarantee that the responses come from a legitimate DNS server. Furthermore, there is no integrity protection, so the data is vulnerable to tampering during transmission. And finally, there is no advanced privacy protection, which means that DNS queries can be observed and analyzed by anyone monitoring the network. Therefore, the total security score is zero. \textit{DNSSEC} does not provide encryption, so the DNS data remains in plaintext and can still be intercepted by unauthorized parties. However, DNSSEC introduces authentication through digital signatures, ensuring that DNS responses originate from a legitimate source. It also provides integrity protection by verifying that the data has not been tampered with. However, it does not provide advanced privacy protection as the DNS queries are not encrypted or obfuscated. Therefore, the total security score is two. \textit{DoT (DNS over TLS)} offers the encryption of DNS queries and responses via TLS and thus ensures that data cannot be read by unauthorized parties. Authentication takes place via TLS certificates, which are used to verify the legitimacy of the communication endpoints. Integrity is ensured as part of the TLS protocol, which prevents tampering with the data during transmission. However, there is no advanced privacy protection beyond encryption, i.e. DNS traffic is not obfuscated beyond what TLS provides. Therefore, the total security score is three. \textit{DoT + DNSSEC} combines the encryption provided by DoT with the additional security measures of DNSSEC. These include TLS-based encryption, which ensures confidentiality, and the authentication and integrity guarantees provided by both TLS certificates and DNSSEC's digital signatures. This combined approach does not provide any additional advanced privacy protection beyond what each individual protocol provides on its own. Therefore, the overall security score remains in the three-point range. \textit{DoH (DNS over HTTPS)} offers comprehensive security with encryption via HTTPS and ensures that DNS queries and responses are encrypted and cannot be intercepted by unauthorized parties. Authentication takes place via HTTPS certificates, which guarantee the legitimacy of the servers involved. The HTTPS protocol also guarantees integrity and thus protects against data tampering. In addition, DoH provides advanced privacy protection by mixing DNS traffic with regular HTTPS web traffic, making it more difficult for observers to distinguish DNS queries from other types of web traffic. Therefore, the overall security rating is four.

\subsubsection{\textbf{Response Time}} We calculate the time required to receive a response from different domain name servers located on different continents by using different forwarder responses in milliseconds. In these experiments, the time required to reach different domain name servers on different continents was calculated using different forwarders and local cache-based responses.  In terms of latency after local caching of DNS responses, traditional DNS and DNSSEC were significantly faster compared to the more secure DNS protocols such as DoT, DoH and DoT + DNSSEC. A summary of the average values can be found in the Table \ref{tab:latency_performance_security}. The introduction of DNSSEC results in a performance loss in latency of approximately 93.2\% compared to traditional DNS and a security score of 2 out of 4. Despite the increase in latency, DNSSEC remains relatively fast compared to other secure protocols. DoT results in a performance loss of 482.5\% and a higher security score of 3 out of 4. However, this security comes at a significant cost in terms of response time, making it less suitable for resource-constrained IoT devices. DoT + DNSSEC brings a performance loss of 537.9\%. While this combination offers comprehensive security features, the additional latency makes it difficult to deploy in scenarios where low response times are critical, such as IoT systems with limited bandwidth. DoH has a performance loss of 606.8\%. Although DoH offers the highest level of security (4 out of 4), its significant latency makes it the least suitable option among the tested protocols for resource-constrained IoT environments. \textit{In summary,} the increased latency can be a substantial drawback, particularly for resource-constrained IoT systems, which often operate with limited bandwidth and require fast, efficient responses. Hence, the trade-off between security and performance must be carefully considered when selecting a DNS protocol for such environments. The results indicate that while enhanced security is crucial, the latency introduced by these protocols may not be acceptable for many IoT applications, where quick response times are essential.


\subsubsection{\textbf{Network Traffic}} 
\begin{figure}[htp!]
\centering
\includegraphics[width=.9\linewidth]{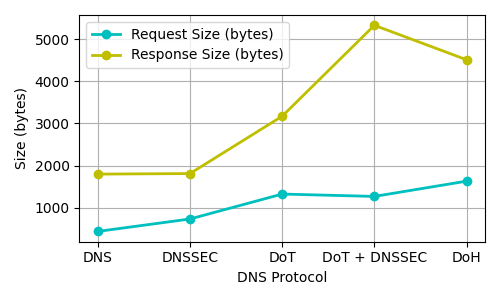}
\vspace{-0.2cm}
\caption{Total Packet Sizes for Request and Response (bytes)} 
\vspace{-0.5cm}
\label{fig:packet}
\end{figure}

In terms of network traffic, we set up Google DNS as a forwarder. We recorded all network traffic and calculated the sum of packets sent and received for each protocol. We recorded the network traffic to determine the differences in packet sizes (in bytes) for each protocol, which are shown in Figure \ref{fig:packet}. According to our observations, DoT+DNSSEC has the highest total overhead when sending requests, while DoH has the highest overhead for requests, and traditional DNS has the lowest in both scenarios. This result was quite expected as DoT has to perform some handshakes beforehand (i.e., 3-way TCP and TLS 1.2 handshakes), and DoH encapsulates the DNS packets into an HTTP frame in addition to DoT's handshakes. Therefore, we believe that most resource-constrained IoT devices will still be able to support such a small number of bytes (i.e., 6,000 bytes at most) of traffic, even for the secure versions of DNS.

\subsection{Discussions on Robustness of Techniques and Recommendations for Future Directions}

Having obtained various DNS results at the edge, both traditional and encrypted/signed DNS, it is important to develop other techniques for resource-constrained devices to ensure robust communication. Table \ref{table:dns_iot_comparison} provides a comparative analysis of DNS protocols for resource-constrained IoT devices, highlighting their trade-offs in terms of security, overhead, privacy, complexity, resource consumption, compatibility, and overall suitability.  Each protocol has its trade-offs, with more secure protocols generally requiring more resources and introducing greater complexity. While traditional DNS over UDP is largely compatible and resource efficient, it lacks security features, making it vulnerable to attacks. DNSSEC increases security through authenticity and integrity checks, but slightly increases overhead and resource consumption. DoT and DoH provide encryption, adding security and privacy, but at the cost of significant overhead and complexity, making them less suitable for highly constrained environments. DoQ attempts to balance security and performance through the QUIC protocol, but remains complex and less compatible. DoC is specifically designed for IoT, offering a good balance between security and efficiency with minimal resource consumption, making it the best option for highly constrained IoT devices.




\begin{table*}[htp!]
\centering
\scriptsize
\caption{Resource-Constrained IoT DNS Protocol Comparison}
\renewcommand{\arraystretch}{1.2} 
\setlength{\tabcolsep}{6pt} 
\begin{tabular}{|l|c|c|c|c|c|c|}
\hline
\rowcolor{gray!30}
\textbf{Feature} & \textbf{DNS} & \textbf{DNSSEC} & \textbf{DoT} & \textbf{DoH} & \textbf{DoQ} & \textbf{DoC} \\
\rowcolor{gray!30}
 & (over UDP) & & (DNS over TLS) & (DNS over HTTPS) & (DNS over QUIC) & (DNS over CoAP) \\
\hline
\textbf{Security} & No encryption or auth. & Authenticity \& integrity & Encryption \& auth. & Encryption \& auth. & Encryption \& auth. & Encryption \\
\hline
\textbf{Overhead} & Lowest & Slightly higher & Higher & Higher & Higher, but potentially & Low (designed for \\
 &  &  &  &  & lower overall & constrained devices) \\
\hline
\textbf{Privacy} & None & Limited & High & High & High & High \\
\hline
\textbf{Complexity} & Simplest & More complex & Moderate & Moderate & More complex & Moderate \\
\hline
\textbf{Resource} & Lowest & Slightly higher & Higher & Higher & Potentially lower & Low \\
\textbf{Consumption} &  &  &  &  & than DoH/DoT &  \\
\hline
\textbf{Compatibility} & Wide & Limited & Increasing & Increasing & Limited & Emerging \\
\hline
\textbf{Suitability for} & Extremely & Some extra & Moderate & Moderate & Promising, but & Very suitable for \\
\textbf{Constrained IoT} & constrained & resources & resources & resources & compatibility concerns & constrained devices \\
\hline
\end{tabular}
\label{table:dns_iot_comparison}
\end{table*}


To ensure robust IoT communication with limited resources, several strategies can be implemented: using lightweight DNS protocols like mDNS or CoAP, which are optimized for low-power devices and constrained networks; deploying DNS resolvers at the network edge to reduce latency and improve response times; leveraging DNS caching on IoT gateways or edge devices to minimize repeated lookups; and setting up multiple DNS servers with failover mechanisms to ensure continuous service availability in case of server failures. In terms of \ac{PQC} transition of these DNS protocols, the level of complexity and resource requirements will significantly affect their adaptability. Traditional DNS over UDP would require significant changes to incorporate PQC due to its lack of encryption and minimal computational requirements, which could negate its simplicity and low resource consumption. DNSSEC, which already includes cryptographic operations for authenticity and integrity, could be extended to support PQC-based algorithms. However, this would further increase computational requirements and overhead, which could limit its use in constrained IoT environments. DoT and DoH, which both rely on TLS and HTTPS for security, would naturally support PQC as part of broader updates to these protocols. However, the increased computational complexity of PQC could exacerbate existing resource and latency issues. DoQ, which is based on the QUIC protocol, offers some flexibility for integrating PQC, especially given QUIC’s efficiency in managing connections, but would still face significant hurdles in resource-constrained environments. Since DoC is specifically designed for IoT), it would need lightweight PQC algorithms to maintain its efficiency, which complicates its upgrade potential but is critical for future-proofing security in IoT deployments. Overall, while all of these protocols could theoretically be upgraded to support PQC, the practical impact on performance, resource consumption and complexity will vary, with simpler protocols experiencing more profound changes.


\section{Future Work}
Based on the findings of this study, we plan to focus on several key areas to enhance the evaluation and applicability of DNS protocols for IoT devices. First, we plan to conduct a comprehensive scalability analysis and measure the respective energy consumption of protocols to evaluate their effects on resource-limited devices under diverse network conditions and device densities. Secondly, we intend to investigate the trade-offs among latency, energy consumption, and memory utilization and develop actionable recommendations tailored for IoT device manufacturers and developers. 
Third, we intend to explore lightweight protocol designs and optimization strategies to improve performance while reducing resource consumption. Furthermore, we plan to investigate the use of Post-Quantum Cryptography (PQC) \cite{hoque2024post, hoque2024exploring, aydeger2024towards} mechanisms to enhance protocol resilience against quantum and classical attacks. 
Finally, our future work will include examining emerging protocols such as DNS-over-QUIC (DoQ) and DNS-over-CoAP (DoC) to assess their appropriateness for IoT applications. These initiatives will establish a thorough framework for choosing DNS protocols suited to particular IoT applications.

\section{Conclusions}
\label{conclusions}

This paper laid the foundation for the implementation of a framework that allows consumers, manufacturers, and providers to make informed decisions about the choice of DNS security protocol depending on their needs and the limitations of the devices. We implemented a DNS server as a \ac{VNF} and tested various DNS protocols, including DNSSEC, DoT and DoH. A number of DNS security protocols were investigated to determine their impact,  reliability, speed, and compatibility for resource-constrained IoT devices. According to the results, DoH has the highest overhead when sending requests, while traditional DNS has the lowest overhead. In terms of response overhead, DoT+DNSSEC causes the highest overhead, while traditional DNS causes the lowest. Before caching, traditional DNS, DNSSEC and DoT had comparable latency, but DoT+DNSSEC and DoH had significantly higher latency. Our results show that we need more efficient methods to enable robust and secure DNS communication for resource-constrained devices.


\ifCLASSOPTIONcaptionsoff
  \newpage
\fi

\balance

\bibliographystyle{ieeetr}
\bibliography{biblio}  

\end{document}